\documentclass[runningheads]{llncs}
\usepackage{graphicx, subfigure}
\usepackage[table,xcdraw]{xcolor}
\usepackage{multirow}
\usepackage{times}
\usepackage{xfp}
\usepackage{amsmath}
\usepackage[normalem]{ulem}
\useunder{\uline}{\ul}{}
\usepackage{hyperref}

\begin{document}
\title{Automation of Citation Screening for Systematic Literature Reviews using Neural Networks: \\A Replicability Study}
\titlerunning{Automation of Citation Screening: a Replicability Study}

\author{Wojciech Kusa\inst{1} \and Allan Hanbury\inst{1,2} \and
Petr Knoth\inst{3, 4}}
\authorrunning{Wojciech Kusa et al.}
\institute{TU Wien, Vienna, Austria,\\
\and
Complexity Science Hub, Vienna, Austria\\
\and
Research Studios Austria, Vienna, Austria\\
\and
The Open University, Milton Keynes, The United Kingdom\\
\email{wojciech.kusa@tuwien.ac.at}
}
\maketitle              % typeset the header of the contribution

\setcounter{footnote}{0} 

\begin{abstract}
In the process of Systematic Literature Review, citation screening is estimated to be one of the most time-consuming steps. Multiple approaches to automate it using various machine learning techniques have been proposed. The first research papers that apply deep neural networks to this problem were published in the last two years. In this work, we conduct a replicability study of the first two deep learning papers for citation screening \cite{Kontonatsios2020UsingReviews,vanDinter2021AProcess} and evaluate their performance on 23 publicly available datasets. While we succeeded in replicating the results of one of the papers, we were unable to replicate the results of the other. We summarise the challenges involved in the replication, including difficulties in obtaining the datasets to match the experimental setup of the original papers and problems with executing the original source code. Motivated by this experience, we subsequently present a simpler model based on averaging word embeddings that outperforms one of the models on 18 out of 23 datasets and is, on average, 72 times faster than the second replicated approach. Finally, we measure the training time and the invariance of the models when exposed to a variety of input features and random initialisations, demonstrating differences in the robustness of these approaches.

\end{abstract}

\section{Introduction}

A systematic literature review is a type of secondary study that summarises all available data fitting pre-specified criteria to answer precise research questions. It uses rigorous scientific methods to minimise bias and generate clear, solid conclusions that health practitioners frequently use to make decisions \cite{Jo2009}.

Unfortunately, conducting systematic reviews is slow, labour intensive and time-consuming as this relies primarily on human effort. A recent estimate shows that conducting a full systematic review takes, on average, 67 weeks \cite{Borah2017AnalysisRegistry}, although another past study reports that the median time to publication was 2.4 years \cite{Tricco2008FollowingStudy}. Furthermore, according to \cite{Shojania2007HowAnalysis}, 23\% of published systematic reviews need updating within two years after completion.

Citation screening (also known as selection of primary studies) is a crucial part of the systematic literature review process \cite{Tsafnat2014SystematicTechnologies}. During this stage, reviewers need to read and comprehend hundreds (or thousands) of documents and decide whether or not they should be included in the systematic review. This decision is made on the basis of comparing each article content with predefined exclusion and inclusion criteria. Traditionally it consists of two stages, the first round of screening titles and abstracts, which is supposed to narrow down the list of potentially relevant items. It is followed by a task appraising the full texts, a more detailed (but also more time-consuming) revision of all included papers from the first stage based on the full text of articles.

Multiple previous studies tried to decrease the completion time of systematic reviews by using text mining methods to semi-automate the citation screening process (see a recent systematic review on this topic: \cite{VanDinter2021}). Using the machine learning paradigm, citation screening could be reduced to a binary classification problem. Then, the task is to train a model using the seed of manually labelled citations that can distinguish between documents to be included (includes) and those to be excluded (excludes). One of the challenges is a significant class imbalance (for 23 benchmark datasets, the maximum percentage of included documents is 27\%, and on average, it is only 7\%). Additionally, existing approaches require training a separate model for each new systematic review.

In this work, we replicate two recent papers related to automated citation screening for systematic literature reviews using neural networks \cite{Kontonatsios2020UsingReviews,vanDinter2021AProcess}. We chose these studies since, to our knowledge, they are the first ones to address this problem using deep neural networks. Both papers represent citation screening as a binary classification task and train an independent model for each dataset. We evaluate the models on 23 publicly available benchmark datasets. We present our challenges regarding replicability in terms of datasets, models and evaluation. In the remaining sections of this article, we will use the name \textbf{Paper A} to refer to the study by Kontonatsios et al. \cite{Kontonatsios2020UsingReviews} and \textbf{Paper B} to indicate work by van Dinter et al. \cite{vanDinter2021AProcess}.

Moreover, we investigate if the models are invariant to different data features and random initialisations. 18 out of 23 datasets are available as a list of Pubmed IDs of the input papers with assigned categories (included or excluded). As we needed to recreate data collection stages for both papers, we wanted to measure if the choice of the document features would influence the final results of the replicated models.

Both papers utilise deep learning due to their claimed substantial superiority over traditional (including shallow neural network) models. We compare the models with previous benchmarks and assess to what extent do these models improve performance over simpler and more traditional models. Finally, we make our data collection and experiment scripts and detailed results publicly available on GitHub\footnote{\url{https://github.com/ProjectDoSSIER/CitationScreeningReplicability}}.

\section{Related Work}

Out of all stages of the systematic review process, the selection of primary studies is known as the most time-consuming step \cite{Bannach-Brown2019MachineError,Sellak2015UsingApproach,Tsafnat2018AutomatedCharacteristics}. It was also automated the most often in the past using text mining methods. According to a recent survey on the topic of automation of systematic literature reviews \cite{VanDinter2021}, 25 out of 41 analysed primary studies published between 2006 and 2020 addressed \mbox{(semi-)automation} of the citation screening process. Another, older systematic review from 2014 found in total 44 studies dealing implicitly or explicitly with the problem of screening workload \cite{OMara-Eves2015}.

Existing approaches to automation of the citation screening process can be categorised into two main groups. The first one uses text classification models \cite{Wallace2010Semi-automatedReviews,Matwin2010AReviews} and the second one screening prioritisation or ranking techniques that exclude items falling below some threshold \cite{Howard2016a,Cohen2008OptimizingPrioritization}. Both groups follow a similar approach. They  train a supervised binary classification algorithm to solve this problem, e.g. Support Vector Machines (SVMs) \cite{Wallace2010Semi-automatedReviews,Cohen2008OptimizingPrioritization}, Naïve Bayes \cite{Matwin2010AReviews} or Random Forest \cite{Khabsa2015LearningInformation}. A significant limitation of these approaches is the need for a large number of human decisions (annotations) that must be completed before developing a reliable model \cite{Tsafnat2018AutomatedCharacteristics}.

Kontonatsios et al. \cite{Kontonatsios2020UsingReviews} (\textbf{Paper A}) was the first one to apply deep learning algorithms to automate the citation screening process. They have used three neural network-based denoising autoencoders to create a feature representation of the documents. This representation was fed into a feed-forward network with a linear SVM classifier trained in a supervised manner to re-order the citations.  Van Dinter et al. \cite{vanDinter2021AProcess} (\textbf{Paper B}) presented the first end-to-end solution to citation screening with a deep neural network. They developed a binary text classification model with the usage of a multi-channel convolutional neural network. Both models claim to yield significant workload savings of at least 10\% on most benchmark review datasets.

A different procedure to automating systematic reviews was presented during the CLEF 2017 eHealth Lab Technology Assisted Reviews in Empirical Medicine task \cite{Kanoulas2017CLEFOverview,Kanoulas2018CLEFOverview}. Here, the user needs to find all relevant documents from a set of PubMed articles given a Boolean query. It overcomes the need for creating an annotated dataset first but makes it harder to incorporate reviewers' feedback.

The recently published BERT model \cite{Devlin2018BERT:Understanding} and its variants have pushed the state of the art for many NLP tasks. Ioannidis \cite{Ioannidis2021AnTask} used BERT-based models to work on document screening within the Technology Assisted Review task achieving better results than the traditional IR baseline models. To our knowledge, this was the first use of a generative neural network model in a document screening task.

\section{Experiment setup}

\subsection{Models}

\subsubsection{DAE-FF}

Paper A presents a neural network-based, supervised feature extraction method combined with a linear Support Vector Machine (SVM) trained to prioritise eligible documents. The data preprocessing pipeline contains stopword removal and stemming with a Porter stemmer. The feature extraction part is implemented as three independent denoising autoencoders (DAE) that learn to reconstruct corrupted Bag-of-Words input vectors. Their concatenated output is used to initialise a supervised feed-forward neural network (FF). These extracted document vectors are subsequently used as an input to an L2-regularised linear SVM classifier. Class imbalance is handled by setting the regularisation parameter $C = 1\times 10^{-6}$.

\subsubsection{Multi-Channel CNN}

Paper B presents a multi-channel convolutional neural network (CNN) to discriminate between includes and excludes. It uses static, pre-trained GloVe word embeddings \cite{Pennington2014GloVe:Representation} to create an input embedding matrix. This embedding is inserted into a series of parallel CNN blocks consisting of a single-dimensional CNN layer followed by global max pooling. Outputs from the layers are concatenated after global pooling and fed into a feed-forward network. The authors experimented with a different number of channels and Conv1D output shapes. Input documents are tokenised and lowercased, punctuation and non-alphabetic tokens are removed. Documents are padded and truncated to a maximum length of 600 tokens. Class imbalance is handled with oversampling. For our replicability study, we have chosen the best performing Model\_2.

\subsubsection{fastText}

We also test a shallow neural network model which is based on fastText word embeddings \cite{Bojanowski2016}. This model is still comparable to more complex deep learning models in many classification tasks. At the same time, it is orders of magnitude faster for training and prediction, making it more suitable for active learning scenarios where reviewers could alter the model's predictions by annotating more documents. To make it even simpler, we do not use pre-trained word embeddings to vectorise documents. Data preprocessing is kept minimal as we only lowercase the text and remove all non-alphanumerical characters.

\subsubsection{Hyperparameters}
Paper A optimised only the number of training epochs for their DAE model. In order to do so, they used two datasets: Statins and BPA reviews and justified this choice with differences between smaller datasets from Clinical and Drug reviews and SWIFT reviews. Other hyperparameters (including the minibatch size and the number of epochs for the feed-forward model) are constant across all datasets. Paper B used the Statins review dataset to tune a set of hyperparameters, including the number of epochs, batch size, dropout, and dense units.

\subsection{Data}

All 23 datasets are summarised in Table \ref{tab:data-statistics}, including the dataset source, number of citations, number and percentage of eligible citations, maximum WSS@95\% score (Section \ref{section:eval}) and the availability of additional bibliographic metadata. Every citation consists of a title, an abstract, and an eligibility label (included or excluded). Moreover, 18 datasets contain also bibliographic metadata. The percentage of eligible citations (includes) varies between datasets, from 0.55\% to 27.04\%, but on average, it is about 7\%, meaning that the datasets are highly imbalanced.

\begin{table}[]

\centering
\caption{Statistics of 23 publicly available datasets used in the experiments on automated citation screening for Systematic Literature Reviews.}
\label{tab:data-statistics}
\resizebox{\textwidth}{!}{%
\begin{tabular}{c|c|c|c|c|c|c|c}
\hline
\textbf{}                 & \textbf{Dataset name}                                   & \textbf{\begin{tabular}[c]{@{}c@{}}Introduced\\ in\end{tabular}}                                                      & \textbf{\# Citations}        & \textbf{\begin{tabular}[c]{@{}c@{}}Included\\ citations\end{tabular}} & \textbf{\begin{tabular}[c]{@{}c@{}}Excluded\\ citations\end{tabular}} & \textbf{\begin{tabular}[c]{@{}c@{}}Maximum \\ WSS@95\%\end{tabular}} & \textbf{\begin{tabular}[c]{@{}c@{}}Bibliographic \\ metadata\end{tabular}} \\ \hline
{\color[HTML]{000000} 1}  & {\color[HTML]{000000} ACEInhibitors}                   & {\color[HTML]{000000} }                                                                                               & {\color[HTML]{000000} 2544}  & {\color[HTML]{000000} 41 (1.6\%)}                                     & {\color[HTML]{000000} 2503 (98.4\%)}                                  & {\color[HTML]{000000} 93.47\%}                                       & {\color[HTML]{000000} Yes}                                                 \\
{\color[HTML]{000000} 2}  & {\color[HTML]{000000} ADHD}                            & {\color[HTML]{000000} }                                                                                               & {\color[HTML]{000000} 851}   & {\color[HTML]{000000} 20 (2.4\%)}                                     & {\color[HTML]{000000} 831 (97.6\%)}                                   & {\color[HTML]{000000} 92.77\%}                                       & {\color[HTML]{000000} Yes}                                                 \\
{\color[HTML]{000000} 3}  & {\color[HTML]{000000} Antihistamines}                  & {\color[HTML]{000000} }                                                                                               & {\color[HTML]{000000} 310}   & {\color[HTML]{000000} 16 (5.2\%)}                                     & {\color[HTML]{000000} 294 (94.8\%)}                                   & {\color[HTML]{000000} 89.84\%}                                       & {\color[HTML]{000000} Yes}                                                 \\
{\color[HTML]{000000} 4}  & {\color[HTML]{000000} Atypical Antipsychotics}         & {\color[HTML]{000000} }                                                                                               & {\color[HTML]{000000} 1120}  & {\color[HTML]{000000} 146 (13.0\%)}                                   & {\color[HTML]{000000} 974 (87.0\%)}                                   & {\color[HTML]{000000} 82.59\%}                                       & {\color[HTML]{000000} Yes}                                                 \\
{\color[HTML]{000000} 5}  & {\color[HTML]{000000} Beta Blockers}                   & {\color[HTML]{000000} }                                                                                               & {\color[HTML]{000000} 2072}  & {\color[HTML]{000000} 42 (2.0\%)}                                     & {\color[HTML]{000000} 2030 (98.0\%)}                                  & {\color[HTML]{000000} 93.07\%}                                       & {\color[HTML]{000000} Yes}                                                 \\
{\color[HTML]{000000} 6}  & {\color[HTML]{000000} Calcium Channel Blockers}        & {\color[HTML]{000000} }                                                                                               & {\color[HTML]{000000} 1218}  & {\color[HTML]{000000} 100 (8.2\%)}                                    & {\color[HTML]{000000} 1118 (91.8\%)}                                  & {\color[HTML]{000000} 87.20\%}                                       & {\color[HTML]{000000} Yes}                                                 \\
{\color[HTML]{000000} 7}  & {\color[HTML]{000000} Estrogens}                       & {\color[HTML]{000000} }                                                                                               & {\color[HTML]{000000} 368}   & {\color[HTML]{000000} 80 (21.7\%)}                                    & {\color[HTML]{000000} 288 (78.3\%)}                                   & {\color[HTML]{000000} 74.35\%}                                       & {\color[HTML]{000000} Yes}                                                 \\
{\color[HTML]{000000} 8}  & {\color[HTML]{000000} NSAIDs}                          & {\color[HTML]{000000} }                                                                                               & {\color[HTML]{000000} 393}   & {\color[HTML]{000000} 41 (10.4\%)}                                    & {\color[HTML]{000000} 352 (89.6\%)}                                   & {\color[HTML]{000000} 85.08\%}                                       & {\color[HTML]{000000} Yes}                                                 \\
{\color[HTML]{000000} 9}  & {\color[HTML]{000000} Opioids}                         & {\color[HTML]{000000} }                                                                                               & {\color[HTML]{000000} 1915}  & {\color[HTML]{000000} 15 (0.8\%)}                                     & {\color[HTML]{000000} 1900 (99.2\%)}                                  & {\color[HTML]{000000} 94.22\%}                                       & {\color[HTML]{000000} Yes}                                                 \\
{\color[HTML]{000000} 10} & {\color[HTML]{000000} Oral Hypoglycemics}              & {\color[HTML]{000000} }                                                                                               & {\color[HTML]{000000} 503}   & {\color[HTML]{000000} 136 (27.0\%)}                                   & {\color[HTML]{000000} 367 (73.0\%)}                                   & {\color[HTML]{000000} 69.16\%}                                       & {\color[HTML]{000000} Yes}                                                 \\
{\color[HTML]{000000} 11} & {\color[HTML]{000000} Proton PumpInhibitors}           & {\color[HTML]{000000} }                                                                                               & {\color[HTML]{000000} 1333}  & {\color[HTML]{000000} 51 (3.8\%)}                                     & {\color[HTML]{000000} 1282 (96.2\%)}                                  & {\color[HTML]{000000} 91.32\%}                                       & {\color[HTML]{000000} Yes}                                                 \\
{\color[HTML]{000000} 12} & {\color[HTML]{000000} Skeletal Muscle Relaxants}       & {\color[HTML]{000000} }                                                                                               & {\color[HTML]{000000} 1643}  & {\color[HTML]{000000} 9 (0.5\%)}                                      & {\color[HTML]{000000} 1634 (99.5\%)}                                  & {\color[HTML]{000000} 94.45\%}                                       & {\color[HTML]{000000} Yes}                                                 \\
{\color[HTML]{000000} 13} & {\color[HTML]{000000} Statins}                         & {\color[HTML]{000000} }                                                                                               & {\color[HTML]{000000} 3465}  & {\color[HTML]{000000} 85 (2.5\%)}                                     & {\color[HTML]{000000} 3380 (97.5\%)}                                  & {\color[HTML]{000000} 92.66\%}                                       & {\color[HTML]{000000} Yes}                                                 \\
{\color[HTML]{000000} 14} & {\color[HTML]{000000} Triptans}                        & {\color[HTML]{000000} }                                                                                               & {\color[HTML]{000000} 671}   & {\color[HTML]{000000} 24 (3.6\%)}                                     & {\color[HTML]{000000} 647 (96.4\%)}                                   & {\color[HTML]{000000} 91.57\%}                                       & {\color[HTML]{000000} Yes}                                                 \\
{\color[HTML]{000000} 15} & {\color[HTML]{000000} Urinary Incontinence}            & \multirow{-15}{*}{{\color[HTML]{000000} \begin{tabular}[c]{@{}c@{}}Drug \\ (Cohen et al., \\ 2006 )\end{tabular}}}    & {\color[HTML]{000000} 327}   & {\color[HTML]{000000} 40 (12.2\%)}                                    & {\color[HTML]{000000} 287 (87.8\%)}                                   & {\color[HTML]{000000} 83.38\%}                                       & {\color[HTML]{000000} Yes}                                                 \\ \hline
\rowcolor[HTML]{EFEFEF} 
{\color[HTML]{000000} }   & {\color[HTML]{000000} Average Drug}           & {\color[HTML]{000000} }                                                                                               & {\color[HTML]{000000} 1249}  & {\color[HTML]{000000} 56 (7.7\%)}                                     & {\color[HTML]{000000} 1192 (92.3\%)}                                  & {\color[HTML]{000000} 87.67\%}                                       & {\color[HTML]{000000} 15/15}                                               \\ \hline
{\color[HTML]{000000} 16} & {\color[HTML]{000000} COPD}                            & {\color[HTML]{000000} }                                                                                               & {\color[HTML]{000000} 1606}  & {\color[HTML]{000000} 196 (12.2\%)}                                   & {\color[HTML]{000000} 1410 (87.8\%)}                                  & {\color[HTML]{000000} 83.36\%}                                       & {\color[HTML]{000000} No}                                                  \\
{\color[HTML]{000000} 17} & {\color[HTML]{000000} Proton Beam}                     & {\color[HTML]{000000} }                                                                                               & {\color[HTML]{000000} 4751}  & {\color[HTML]{000000} 243 (5.1\%)}                                    & {\color[HTML]{000000} 4508 (94.9\%)}                                  & {\color[HTML]{000000} 90.14\%}                                       & {\color[HTML]{000000} No}                                                  \\
{\color[HTML]{000000} 18} & {\color[HTML]{000000} Micro Nutrients}                 & \multirow{-3}{*}{{\color[HTML]{000000} \begin{tabular}[c]{@{}c@{}}Clinical\\ (Wallace et al., \\ 2010)\end{tabular}}} & {\color[HTML]{000000} 4010}  & {\color[HTML]{000000} 258 (6.4\%)}                                    & {\color[HTML]{000000} 3752 (93.6\%)}                                  & {\color[HTML]{000000} 88.87\%}                                       & {\color[HTML]{000000} No}                                                  \\ \hline
\rowcolor[HTML]{EFEFEF} 
{\color[HTML]{000000} }   & {\color[HTML]{000000} Average Clinical}       & {\color[HTML]{000000} }                                                                                               & {\color[HTML]{000000} 3456}  & {\color[HTML]{000000} 232 (7.9\%)}                                    & {\color[HTML]{000000} 3223 (92.1\%)}                                  & {\color[HTML]{000000} 87.45\%}                                       & {\color[HTML]{000000} 0/3}                                                 \\ \hline
{\color[HTML]{000000} 19} & {\color[HTML]{000000} PFOA/PFOS}                       & {\color[HTML]{000000} }                                                                                               & {\color[HTML]{000000} 6331}  & {\color[HTML]{000000} 95 (1.5\%)}                                     & {\color[HTML]{000000} 6236 (98.5\%)}                                  & {\color[HTML]{000000} 93.56\%}                                       & {\color[HTML]{000000} Yes}                                                 \\
{\color[HTML]{000000} 20} & {\color[HTML]{000000} Bisphenol A (BPA)}               & {\color[HTML]{000000} }                                                                                               & {\color[HTML]{000000} 7700}  & {\color[HTML]{000000} 111 (1.4\%)}                                    & {\color[HTML]{000000} 7589 (98.6\%)}                                  & {\color[HTML]{000000} 93.62\%}                                       & {\color[HTML]{000000} Yes}                                                 \\
{\color[HTML]{000000} 21} & {\color[HTML]{000000} Transgenerational}               & {\color[HTML]{000000} }                                                                                               & {\color[HTML]{000000} 48638} & {\color[HTML]{000000} 765 (1.6\%)}                                    & {\color[HTML]{000000} 47873 (98.4\%)}                                 & {\color[HTML]{000000} 93.51\%}                                       & {\color[HTML]{000000} Yes}                                                 \\
{\color[HTML]{000000} 22} & {\color[HTML]{000000} Fluoride and neurotoxicity}      & {\color[HTML]{000000} }                                                                                               & {\color[HTML]{000000} 4479}  & {\color[HTML]{000000} 51 (1.1\%)}                                     & {\color[HTML]{000000} 4428 (98.9\%)}                                  & {\color[HTML]{000000} 93.91\%}                                       & {\color[HTML]{000000} No}                                                  \\
{\color[HTML]{000000} 23} & {\color[HTML]{000000} Neuropathic pain | CAMRADES}     & \multirow{-5}{*}{{\color[HTML]{000000} \begin{tabular}[c]{@{}c@{}}SWIFT \\ (Howard et al., \\ 2016)\end{tabular}}}    & {\color[HTML]{000000} 29207} & {\color[HTML]{000000} 5011 (17.2\%)}                                  & {\color[HTML]{000000} 24196 (82.8\%)}                                 & {\color[HTML]{000000} 78.70\%}                                       & {\color[HTML]{000000} No}                                                  \\ \hline
\rowcolor[HTML]{EFEFEF} 
{\color[HTML]{000000} }   & {\color[HTML]{000000} Average SWIFT}          & {\color[HTML]{000000} }                                                                                               & {\color[HTML]{000000} 19271} & {\color[HTML]{000000} 1206 (4.6\%)}                                   & {\color[HTML]{000000} 18064 (95.4\%)}                                 & {\color[HTML]{000000} 90.66\%}                                       & {\color[HTML]{000000} 3/5}                                                 \\ \hline
\rowcolor[HTML]{D3D3D3} 
{\color[HTML]{000000} }   & {\color[HTML]{000000} Average (All datasets)} & {\color[HTML]{000000} }                                                                                               & {\color[HTML]{000000} 5454}  & {\color[HTML]{000000} 329 (7.0\%)}                                    & {\color[HTML]{000000} 5125 (93.0\%)}                                  & {\color[HTML]{000000} 88.29\%}                                       & {\color[HTML]{000000} 18/23}                                               \\ \hline
\end{tabular}%
\vspace{-.3cm}

}

\end{table}

Cohen et al. \cite{Cohen2006} was the first one to introduce datasets for training and evaluation of citation screening. They constructed a test collection for 15 different systematic review topics produced by the Oregon Evidence-based Practice Centre (EPC) related to the efficacy of medications in several drug classes.

Another three datasets for evaluation of automated citation screening were released by Wallace et al. \cite{Wallace2010Semi-automatedReviews}. These systematic reviews are related to the clinical outcomes of various treatments. Both drug and clinical reviews contain a small number of citations (varying from 310 to 4751).

The third group of datasets was introduced by Howard et al. \cite{Howard2016a} and consists of five substantially larger reviews (from 4479 to 48 638 citations) that have been used to assess the performance of the SWIFT-review tool. They were created using broader search strategies which justifies a higher number of citations.

Paper A trained and evaluated their model on all 23 datasets coming from three categories. Paper B used 20 datasets from the Clinical and SWIFT categories. Paper B states that, on average, 5.2\% of abstracts are missing in all 20 datasets, varying between 0\% for \textit{Neuropathic Pain} and 20.82\% for \textit{Statins}. Compared to previous papers, Paper B reports fewer citations for three datasets (Table 6 in the original paper): \textit{Statins}, \textit{PFOA/PFOS} and \textit{Neuropathic Pain}. This difference is insignificant compared to the dataset size, e.g. 29207 versus 29202 for \textit{Neuropathic Pain}, so it should not influence the model evaluation.

\subsection{Evaluation} 
\label{section:eval}

Evaluation of automated citation screening can be very challenging. Traditional metrics used for classification tasks like precision, recall, or F-score cannot capture what we intend to measure in this task. For an automated system to be beneficial to systematic reviewers, it should save time and miss as few relevant papers as possible. Previous studies suggested that recall should not be lower than 95\%, and at the same time, precision should be as high as possible \cite{Cohen2006}.

\subsubsection{Work saved over sampling} at r\% recall (WSS@r\%) is a primary metric for evaluation of automated citation screening. It was first introduced and described by Cohen et al. \cite{Cohen2006} as ``the percentage of papers that meet the original search criteria that the reviewers do not have to read (because they have been screened out by the classifier).'' It estimates the human screening workload reduction by using automation tools, assuming a fixed recall level of r\%. WSS@r\%, given a recall of r\%, is defined as follows:

\begin{equation*}
  WSS@r\% = \frac{TN + FN}{N} - \left(1 - r\right)
\end{equation*}

\noindent where TN is the number of true negatives, FN is the number of false negatives, and N is the total number of documents. Based on previous studies, we fix the recall at 95\% and compute the WSS@95\% score.

One drawback of this metric described by \cite{Cohen2006} is that it does not take into account time differences caused by varying lengths of documents and also the time needed to review a full-text article compared to only reading the title and the abstract. 

A further drawback of WSS is that the maximum WSS value depends on the ratio of included/excluded samples. A perfectly balanced dataset can achieve a maximum value of WSS@95\% = 0.45, whereas a highly imbalanced dataset with a 5\%/95\% split can obtain a maximum WSS@95\% score of 0.9. Consequently, it does not make sense to compare the results nor average them across different datasets (as done in Paper A and B).

For our replicability study, we decided to use the implementations of the WSS metric provided by Papers A and B.

\subsubsection{Cross-validation} 

Both papers use a stratified $10\times2$ cross-validation for evaluation. In this setting, data is randomly split in half: one part is used to train the classifier, and the other is left for testing. This process is then repeated ten times, and the results are accumulated from all ten runs. We also use this approach to evaluate the quality of all three models.

\subsection{Code}

The authors of both papers uploaded their code into public GitHub repositories: \footnote{\url{https://github.com/gkontonatsios/DAE-FF}}\textsuperscript{,}\footnote{\url{https://github.com/rvdinter/multichannel-cnn-citation-screening}}. Both models were written in Python 3 and depend primarily on TensorFlow and Keras deep learning frameworks \cite{Abadi2015TensorFlow:Systems}. The whole implementation was uploaded in four commits for Paper A and one for Paper B (excluding commits containing only documentation). Except for the code, there is no information about versions of the packages used to train and evaluate the models. This missing information is crucial for replicability, as, for TensorFlow alone, in 2020, there were 27 different releases related to 6 different MINOR versions\footnote{\url{https://pypi.org/project/tensorflow/\#history}}.

The model prepared by Paper B uses also pre-trained 100-dimensional GloVe word embeddings which we downloaded separately from the original authors' website\footnote{\url{https://nlp.stanford.edu/data/glove.6B.zip}} according to the instructions provided by the Paper B GitHub Readme.

Both papers did not include the original datasets they used to train and evaluate their models. Paper A provided sample data consisting of 100 documents which presents the input data format accepted by their model, making it easier to re-run the experiments. Paper B does not include sample data but describes where and how to collect and process the datasets.

\section{Results and discussion}

\subsection{Replicability study}

WSS@95\% scores from older benchmarks and original papers, along with our replicated results, are presented in Table \ref{tab:wss95-results}. For all datasets, both Paper A and B provide only mean WSS@95\% score from cross-validation runs. Therefore, we were not able to measure statistical significance between our replicated results and the original ones. To quantify the difference, we decided to calculate the absolute delta between reported and replicated scores: $ |x - y|$. Both models report a random seed for the cross-validation splits but not for the model optimisation. Usage of different seeds for model optimisation might be one of the reasons why we were not able to achieve the same results.

% \usepackage{graphicx}
% \usepackage[normalem]{ulem}
\begin{table}[!htb]
\centering
\caption{WSS@95\% results for replicated models compared with original results and benchmark models. WSS@95\% scores are averages across ten validation runs for each of the 23 review datasets. \uline{Underlined} scores indicate the highest score within the three tested models, \textbf{bold} values indicate the highest score overall.}
\label{tab:wss95-results}
% \setlength{}{}
% \tabcolsep{}
\resizebox{\textwidth}{!}{%
\begin{tabular}{c|cccc|ccc|ccc|c}
\hline
Dataset name &
  \begin{tabular}[c]{@{}c@{}}Cohen \\ (2006)\end{tabular} &
  \begin{tabular}[c]{@{}c@{}}Matwin \\ (2010)\end{tabular} &
  \begin{tabular}[c]{@{}c@{}}Cohen \\ (2008/\\ 2011)\end{tabular} &
  \begin{tabular}[c]{@{}c@{}}Howard \\ (2016)\end{tabular} &
  \textbf{Paper A} &
  \begin{tabular}[c]{@{}c@{}}Paper A\\ replicated\end{tabular} &
  \begin{tabular}[c]{@{}c@{}}Absolute\\ delta\end{tabular} &
  \textbf{Paper B} &
  \begin{tabular}[c]{@{}c@{}}Paper B\\ replicated\end{tabular} &
  \begin{tabular}[c]{@{}c@{}}Absolute\\ delta\end{tabular} &
  \begin{tabular}[c]{@{}c@{}}fastText\\ classifier\end{tabular} \\ \hline
ACEInhibitors &
  .566 &
  .523 &
  .733 &
  \textbf{.801} &
  {\ul .787} &
  .785 &
  0.16\% &
  .783 &
  .367 &
  41.59\% &
  .783 \\
ADHD &
  .680 &
  .622 &
  .526 &
  \textbf{.793} &
  .665 &
  .639 &
  2.58\% &
  .698 &
  {\ul .704} &
  0.57\% &
  .424 \\
Antihistamines &
  .000 &
  .149 &
  .236 &
  .137 &
  {\ul \textbf{.310}} &
  .275 &
  3.48\% &
  .168 &
  .135 &
  3.32\% &
  .047 \\
Atypical Antipsychotics &
  .141 &
  .206 &
  .170 &
  .251 &
  {\ul \textbf{.329}} &
  .190 &
  13.92\% &
  .212 &
  .081 &
  13.15\% &
  .218 \\
Beta Blockers &
  .284 &
  .367 &
  .465 &
  .428 &
  {\ul \textbf{.587}} &
  .462 &
  12.52\% &
  .504 &
  .399 &
  10.51\% &
  .419 \\
Calcium Channel Blockers &
  .122 &
  .234 &
  .430 &
  \textbf{.448} &
  {\ul .424} &
  .347 &
  7.66\% &
  .159 &
  .069 &
  9.03\% &
  .178 \\
Estrogens &
  .183 &
  .375 &
  .414 &
  \textbf{.471} &
  {\ul .397} &
  .369 &
  2.80\% &
  .119 &
  .083 &
  3.56\% &
  .306 \\
NSAIDs &
  .497 &
  .528 &
  .672 &
  \textbf{.730} &
  {\ul .723} &
  .735 &
  1.18\% &
  .571 &
  .601 &
  2.98\% &
  .620 \\
Opioids &
  .133 &
  .554 &
  .364 &
  \textbf{.826} &
  .533 &
  {\ul .580} &
  4.71\% &
  .295 &
  .249 &
  4.58\% &
  .559 \\
Oral Hypoglycemics &
  .090 &
  .085 &
  \textbf{.136} &
  .117 &
  .095 &
  {\ul .123} &
  2.80\% &
  .065 &
  .013 &
  5.21\% &
  .098 \\
Proton PumpInhibitors &
  .277 &
  .229 &
  .328 &
  .378 &
  {\ul\textbf{.400}} &
  .299 &
  10.13\% &
  .243 &
  .129 &
  11.38\% &
  .283 \\
Skeletal Muscle Relaxants &
  .000 &
  .265 &
  .374 &
  \textbf{.556} &
  .286 &
  .286 &
  0.04\% &
  .229 &
  {\ul .300} &
  7.14\% &
  .090 \\
Statins &
  .247 &
  .315 &
  .491 &
  .435 &
  {\ul \textbf{.566}} &
  .487 &
  7.93\% &
  .443 &
  .283 &
  16.03\% &
  .409 \\
Triptans &
  .034 &
  .274 &
  .346 &
  .412 &
  .434 &
  .412 &
  2.24\% &
  .266 &
  {\ul \textbf{.440}} &
  17.38\% &
  .210 \\
Urinary Incontinence &
  .261 &
  .296 &
  .432 &
  \textbf{.531} &
  {\ul \textbf{.531}} &
  .483 &
  4.81\% &
  .272 &
  .180 &
  9.21\% &
  .439 \\ \hline
  \rowcolor[HTML]{EFEFEF}
Average Drug &
  .234 &
  .335 &
  .408 &
  \textbf{.488} &
  {\ul .471} &
  .431 &
  5.13\% &
  .335 &
  .269 &
  10.37\% &
  .339 \\ \hline
COPD &
  — &
  — &
  — &
  — &
  {\ul \textbf{.666}} &
  .665 &
  0.07\% &
  — &
  .128 &
  — &
  .312 \\
Proton Beam &
  — &
  — &
  — &
  — &
  {\ul \textbf{.816}} &
  .812 &
  0.39\% &
  — &
  .357 &
  — &
  .733 \\
Micro Nutrients &
  — &
  — &
  — &
  — &
  .662 &
  {\ul \textbf{.663}} &
  0.08\% &
  — &
  .199 &
  — &
  .608 \\ \hline
  \rowcolor[HTML]{EFEFEF}
Average Clinical &
  — &
  — &
  — &
  — &
  {\ul \textbf{.715}} &
  .713 &
  0.18\% &
  — &
  .228 &
  — &
  .551 \\ \hline
PFOA/PFOS &
  — &
  — &
  — &
  .805 &
  {\ul \textbf{.848}} &
  .838 &
  0.97\% &
  .071 &
  .305 &
  23.44\% &
  .779 \\
Bisphenol A (BPA) &
  — &
  — &
  — &
  .752 &
  {\ul \textbf{.793}} &
  .780 &
  1.34\% &
  .792 &
  .369 &
  42.31\% &
  .637 \\
Transgenerational &
  — &
  — &
  — &
  .714 &
  .707 &
  {\ul \textbf{.718}} &
  1.14\% &
  .708 &
  .000 &
  70.80\% &
  .368 \\
Fluoride and neurotoxicity &
  — &
  — &
  — &
  .870 &
  .799 &
  .806 &
  0.68\% &
  {\ul \textbf{.883}} &
  .808 &
  7.48\% &
  .390 \\
Neuropathic pain &
  — &
  — &
  — &
  \textbf{.691} &
  .608 &
  .598 &
  1.03\% &
  {\ul .620} &
  .091 &
  52.89\% &
  .613 \\ \hline
  \rowcolor[HTML]{EFEFEF}
Average SWIFT &
  — &
  — &
  — &
  \textbf{.766} &
  {\ul .751} &
  .748 &
  1.03\% &
  .615 &
  .315 &
  39.38\% &
  .557 \\ \hline
  \rowcolor[HTML]{CCCCCC}
Average (all datasets) &
  — &
  — &
  — &
  — &
  {\ul \textbf{.564}} &
  .537 &
  3.59\% &
  — &
  .273 &
  17.63\% &
  .414 \\ \hline
\end{tabular}%
}

\end{table}

For two datasets (\textit{Bisphenol A (BPA)} and \textit{Triptans}), Paper A reports two different results for the DAE-FF model (Tables 5 and 6 in the original paper). We suppose this was only a typing mistake, as we managed to infer the actual values based on the averaged WSS@95\% score from all datasets available in the original paper.

The average delta between our replicated results and the original ones from Paper A is 3.59\%. Only for three datasets is this value higher than 10\%. If we consider different seeds used for training models, these results confirm the successful replication of Paper A's work.

For Paper B, the average delta is 17.63\%. For 10 out of 20 datasets, this delta is more than 10\%. For the two largest datasets: \textit{Transgenerational} and \textit{Neuropathic Pain} we were not able to successfully train the Multi-Channel CNN model. All of these results raise concerns about replicability. 

Paper B also tried to replicate the DAE-FF model from Paper A. They stated that \textit{``(...) we aimed to replicate the model (...) with open-source code via GitHub. However, we could not achieve the same scores using our dataset. After emailing the primary author, we were informed that he does not have access to his datasets anymore, which means their study cannot be fully replicated.''}. Our results are contrary to findings by Paper B: we managed to replicate the results of Paper A successfully without having access to their original datasets. Unfortunately, Paper B does not present any quantitative results of their replicability study. Therefore, we cannot draw any conclusions regarding those results as we do not know what Paper B authors meant by \textit{``cannot be fully replicated''}.

\begin{figure*}[!tbh]

    \centering
    \subfigure[ ADHD review dataset.]{\includegraphics[width=0.49\linewidth]{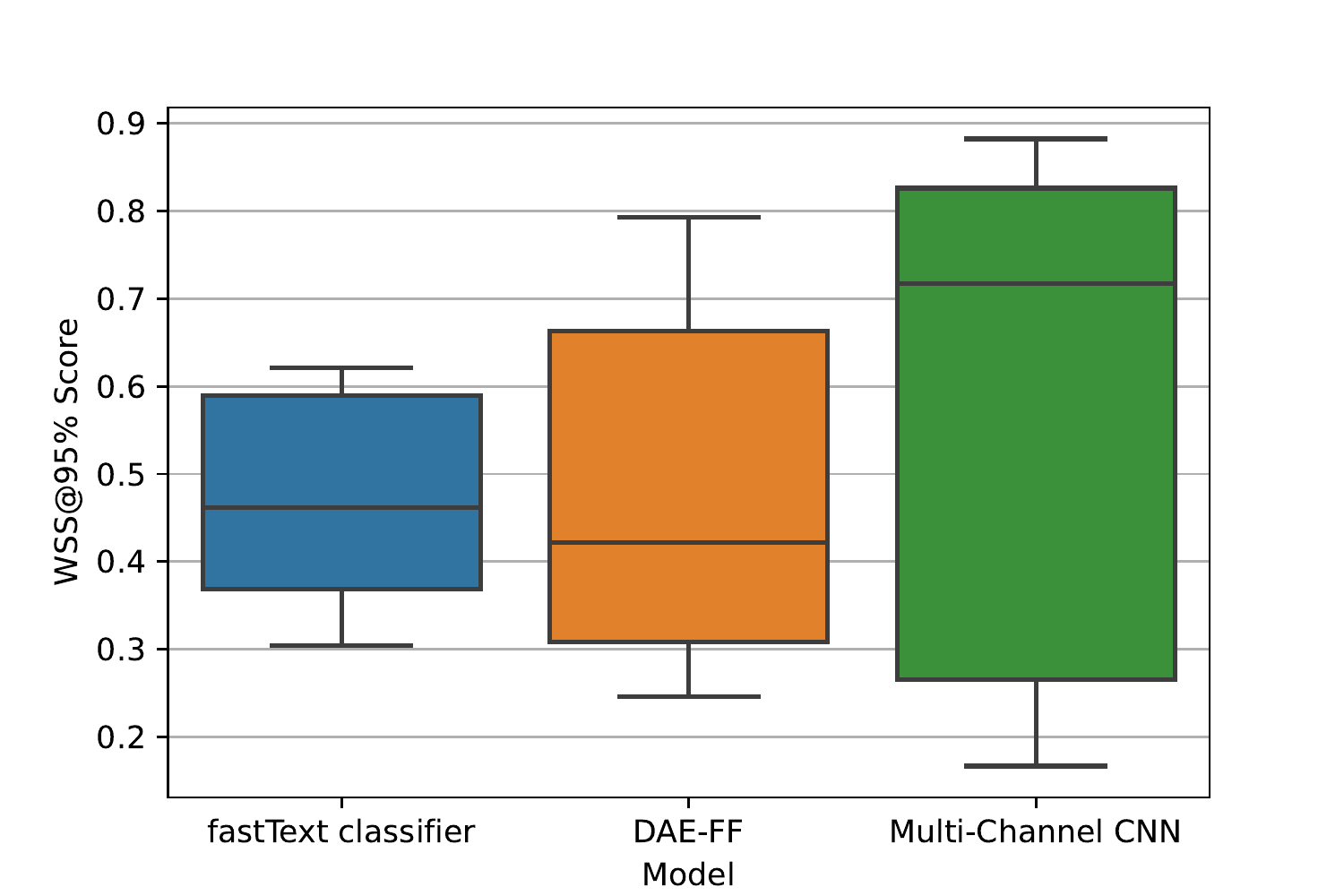}} %
    \subfigure[ Proton Beam review dataset.]{\includegraphics[width=0.49\linewidth]{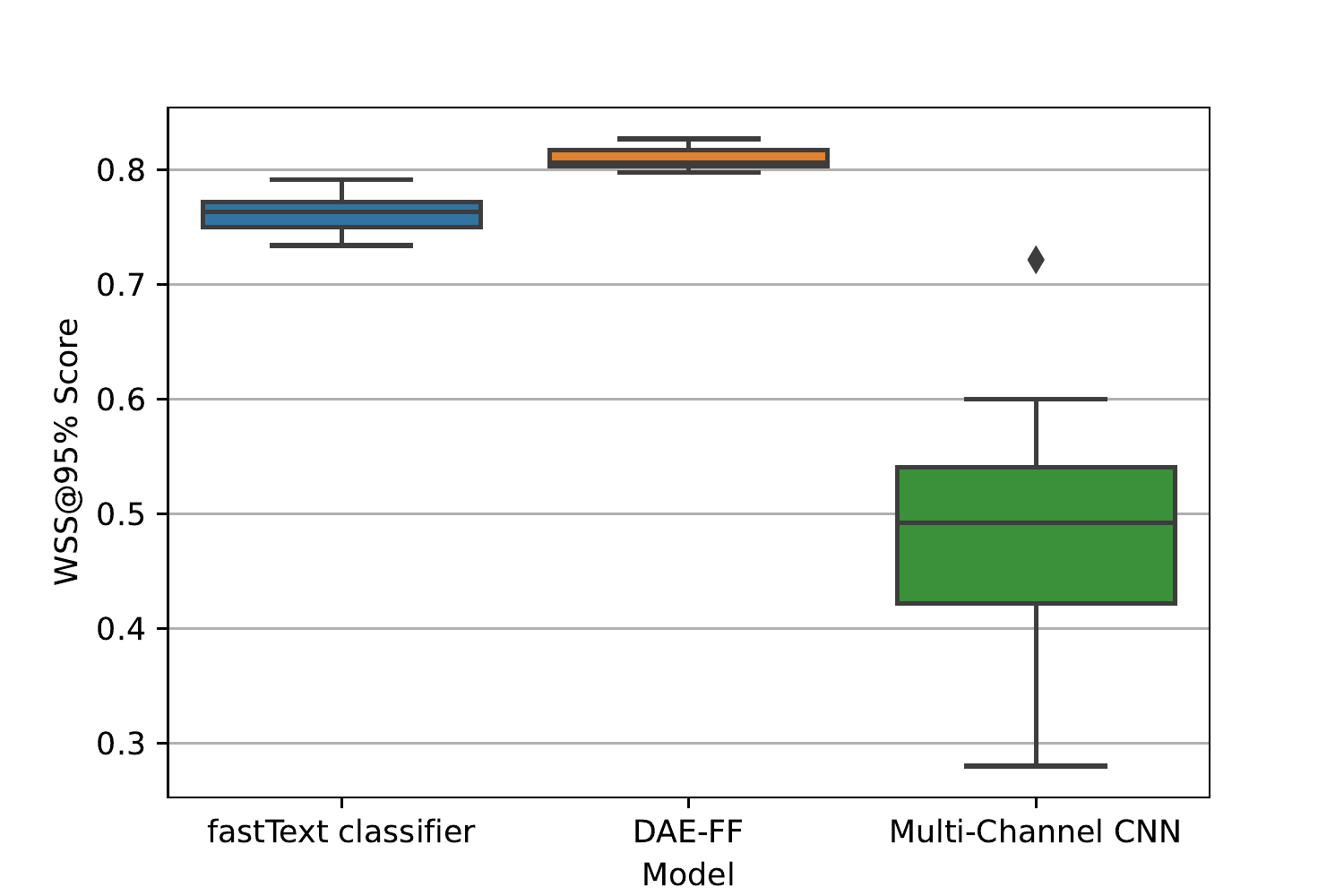}} %
    \caption{Example boxplots with WSS@95\% scores for three models. Input features are titles and abstracts.}%
    \label{fig:score_dist}

\end{figure*}

Figure \ref{fig:score_dist} presents results for \textit{ADHD} and \textit{Proton Beam} datasets for all three models. The Multi-Channel CNN model has the widest range of WSS@95\% scores across cross-validation runs. This is especially evident on the datasets from the Clinical group (i.e. \textit{Proton Beam}), for which the DAE-FF and fastText models yield very steady results across every cross-validation fold. This could mean that the Multi-Channel CNN model is less stable, and its good performance is dependant on random initialisation.

Next, we compare our replicated results and the original ones from Paper A and B to previous benchmark studies. Paper A only compares their model to custom baseline methods and does not mention the previous state of the art results. None of the tested neural network-based models can improve on the results by Howard et al. \cite{Howard2016a}, which uses a log-linear model with word-score and topic-weight features to classify the citations. This means that even though deep neural network models can provide significant gains in WSS@95\% scores, they can still be outperformed by classic statistical methods.

\subsection{Impact of input features}

As we encountered memory problems when training the Paper B model on \textit{Transgenerational} and \textit{Neuropathic pain} datasets, we exclude these two datasets from our comparisons in the remaining experiments.

None of the papers provided the original input data used to train the models. We wanted to measure if the results depend on how that input data was gathered. We implemented two independent data gathering scripts using the biopython package as suggested by Paper B to obtain 18 out of 23 datasets. One implementation relied on the Medline module, where a document was represented as a dictionary of all available fields. The second implementation returned all possible fields (title, abstract, author and journal information) concatenated in a single string. Furthermore, we examined how robust the models are, if the input data contained only titles or abstracts of the citations. Results are presented in the Table \ref{tab:document-features}.

% \usepackage{graphicx}
% \usepackage[normalem]{ulem}
\begin{table}[!htb]
\centering
\caption{Influence of input document features on the WSS@95\% score for three tested models. ``All features'' means a single string containing all possible fields. For each row, \textbf{bold} values indicate the highest score for each model, \uline{underlined} scores are best overall.}
\label{tab:document-features}
\resizebox{\textwidth}{!}{%
\begin{tabular}{c|cccc|cccc|cccc}
\hline
\multicolumn{1}{l|}{}                                               & \multicolumn{4}{c|}{DAE-FF}                                                                                                                                                                                                               & \multicolumn{4}{c|}{Multi-Channel CNN}                                                                                                                                                                                                    & \multicolumn{4}{c}{fastText classifier}                                                                                                                                                                                                  \\ \hline
\multicolumn{1}{c|}{Dataset name}                                    & \begin{tabular}[c]{@{}c@{}}All \\ features\end{tabular} & \begin{tabular}[c]{@{}c@{}}Title and\\ Abstract\end{tabular} & \begin{tabular}[c]{@{}c@{}}Abstract \\ only\end{tabular} & \begin{tabular}[c]{@{}c@{}}Title \\ only\end{tabular} & \begin{tabular}[c]{@{}c@{}}All \\ features\end{tabular} & \begin{tabular}[c]{@{}c@{}}Title and\\ Abstract\end{tabular} & \begin{tabular}[c]{@{}c@{}}Abstract \\ only\end{tabular} & \begin{tabular}[c]{@{}c@{}}Title \\ only\end{tabular} & \begin{tabular}[c]{@{}c@{}}All \\ features\end{tabular} & \begin{tabular}[c]{@{}c@{}}Title and\\ Abstract\end{tabular} & \begin{tabular}[c]{@{}c@{}}Abstract \\ only\end{tabular} & \begin{tabular}[c]{@{}c@{}}Title \\ only\end{tabular} \\ \hline
ACEInhibitors                                                       & .785                                                   & .709                                                        & .658                                                    & {\ul \textbf{.806}}                                  & .367                                                   & .461                                                        & .648                                                    & .525                                                 & \textbf{.783}                                          & .776                                                        & .765                                                    & .441                                                 \\
ADHD                                                                & .639                                                   & .500                                                        & .404                                                    & \textbf{.651}                                        & {\ul \textbf{.704}}                                    & .528                                                        & .692                                                    & .580                                                 & .424                                                   & \textbf{.470}                                               & .444                                                    & .200                                                 \\
Antihistamines                                                      & {\ul \textbf{.275}}                                    & .168                                                        & .265                                                    & .016                                                 & .135                                                   & \textbf{.204}                                               & .114                                                    & .105                                                 & .047                                                   & .124                                                        & .175                                                    & \textbf{.192}                                        \\
Atypical Antipsychotics                                             & .190                                                   & .221                                                        & {\ul \textbf{.230}}                                     & .046                                                 & .081                                                   & \textbf{.086}                                               & .050                                                    & .013                                                 & \textbf{.218}                                          & .188                                                        & .185                                                    & .095                                                 \\
Beta Blockers                                                       & {\ul \textbf{.462}}                                    & .451                                                        & .390                                                    & .408                                                 & \textbf{.399}                                          & .243                                                        & .134                                                    & .211                                                 & \textbf{.419}                                          & \textbf{.419}                                               & .407                                                    & .262                                                 \\
Calcium Channel Blockers                                            & {\ul \textbf{.347}}                                    & .337                                                        & .297                                                    & .137                                                 & .069                                                   & .083                                                        & .004                                                    & \textbf{.117}                                        & .178                                                   & .139                                                        & .060                                                    & \textbf{.244}                                        \\
Estrogens                                                           & {\ul \textbf{.369}}                                    & .358                                                        & .331                                                    & .145                                                 & .083                                                   & .076                                                        & .051                                                    & \textbf{.092}                                        & \textbf{.306}                                          & .199                                                        & .108                                                    & .241                                                 \\
NSAIDs                                                              & {\ul \textbf{.735}}                                    & .679                                                        & .690                                                    & .658                                                 & \textbf{.601}                                          & .443                                                        & .358                                                    & .225                                                 & \textbf{.620}                                          & .506                                                        & .512                                                    & .535                                                 \\
Opioids                                                             & {\ul \textbf{.580}}                                    & .513                                                        & .499                                                    & .280                                                 & .249                                                   & \textbf{.420}                                               & .413                                                    & .287                                                 & \textbf{.559}                                          & .558                                                        & .534                                                    & .245                                                 \\
Oral Hypoglycemics                                                  & .123                                                   & {\ul \textbf{.129}}                                         & .107                                                    & .019                                                 & .013                                                   & \textbf{.021}                                               & .004                                                    & .005                                                 & \textbf{.098}                                          & .049                                                        & .042                                                    & .016                                                 \\
Proton PumpInhibitors                                               & \textbf{.299}                                          & .291                                                        & .153                                                    & .285                                                 & \textbf{.129}                                          & .121                                                        & .059                                                    & .118                                                 & .283                                                   & .228                                                        & .174                                                    & {\ul \textbf{.360}}                                  \\
Skeletal Muscle Relaxants                                           & .286                                                   & .327                                                        & {\ul \textbf{.430}}                                     & .125                                                 & .300                                                   & \textbf{.329}                                               & .242                                                    & .202                                                 & .090                                                   & .142                                                        & .180                                                    & \textbf{.210}                                        \\
Statins                                                             & {\ul \textbf{.487}}                                    & .434                                                        & .392                                                    & .255                                                 & \textbf{.283}                                          & .231                                                        & .120                                                    & .082                                                 & \textbf{.409}                                          & .376                                                        & .281                                                    & .228                                                 \\
Triptans                                                            & \textbf{.412}                                          & .253                                                        & .320                                                    & .199                                                 & {\ul \textbf{.440}}                                    & .404                                                        & .407                                                    & .129                                                 & .210                                                   & .205                                                        & \textbf{.211}                                                    & .075                                        \\
Urinary Incontinence                                                & .483                                                   & {\ul \textbf{.531}}                                         & .482                                                    & .372                                                 & \textbf{.180}                                          & .161                                                        & .046                                                    & .099                                                 & \textbf{.439}                                          & .310                                                        & .170                                                    & .434                                                 \\ \hline
\rowcolor[HTML]{EFEFEF} 
Average Drug                                                        & {\ul \textbf{.431}}                                    & .394                                                        & .373                                                    & .293                                                 & \textbf{.269}                                          & .254                                                        & .223                                                    & .185                                                 & \textbf{.339}                                          & .313                                                        & .283                                                    & .252                                                 \\ \hline
COPD                                                                & .665                                                   & .665                                                        & .676                                                    & {\ul \textbf{.677}}                                  & .128                                                   & \textbf{.372}                                               & .087                                                    & .093                                                 & .312                                                   & \textbf{.553}                                               & .546                                                    & .545                                                 \\
Proton Beam                                                         & {\ul \textbf{.812}}                                    & .810                                                        & .790                                                    & .799                                                 & .357                                                   & .489                                                        & .408                                                    & \textbf{.559}                                        & .733                                                   & .761                                                        & \textbf{.771}                                           & \textbf{.771}                                        \\
Micro Nutrients                                                     & .663                                                   & .648                                                        & .665                                                    & {\ul \textbf{.677}}                                  & .199                                                   & .255                                                        & .251                                                    & \textbf{.268}                                        & \textbf{.608}                                          & .602                                                        & .605                                                    & .601                                                 \\ \hline
\rowcolor[HTML]{EFEFEF} 
Average Clinical                                                    & \textbf{.713}                                          & .708                                                        & .670                                                    & {\ul \textbf{.718}}                                  & .228                                                   & \textbf{.372}                                               & .249                                                    & .307                                                 & .551                                                   & .638                                                        & \textbf{.640}                                           & .639                                                 \\ \hline
PFOA/PFOS                                                           & .713                                                   & .839                                                        & {\ul \textbf{.847}}                                     & .696                                                 & .305                                                   & \textbf{.405}                                               & .391                                                    & .109                                                 & .779                                                   & \textbf{.796}                                               & .778                                                    & .292                                                 \\
Bisphenol A (BPA)                                                   & {\ul \textbf{.780}}                                    & .754                                                        & .715                                                    & .631                                                 & .369                                                   & .300                                                        & \textbf{.612}                                           & .182                                                 & \textbf{.637}                                          & .630                                                        & .499                                                    & .079                                                 \\
Fluoride and neurotoxicity                                          & .806                                                   & {\ul \textbf{.838}}                                         & .758                                                    & .726                                                 & \textbf{.808}                                          & .688                                                        & .654                                                    & .452                                                 & \textbf{.390}                                          & .375                                                        & .292                                                    & .250                                                 \\
\rowcolor[HTML]{EFEFEF} 
Average SWIFT                                                       & .766                                                   & {\ul \textbf{.782}}                                         & .774                                                    & .684                                                 & .494                                                   & .464                                                        & \textbf{.552}                                           & .247                                                 & \textbf{.602}                                          & .600                                                        & .523                                                    & .207                                                 \\
\rowcolor[HTML]{CCCCCC} 
\multicolumn{1}{c|}{\cellcolor[HTML]{CCCCCC}Average (All datasets)} & {\ul \textbf{.520}}                                    & .498                                                        & .481                                                    & .410                                                 & .295                                                   & \textbf{.301}                                               & .274                                                    & .212                                                 & \textbf{.407}                                          & .400                                                        & .368                                                    & .301                                                 \\ \hline
\end{tabular}%
}

\end{table}

\begin{figure}[!thb]
\centering
  \includegraphics[width=0.6\linewidth]{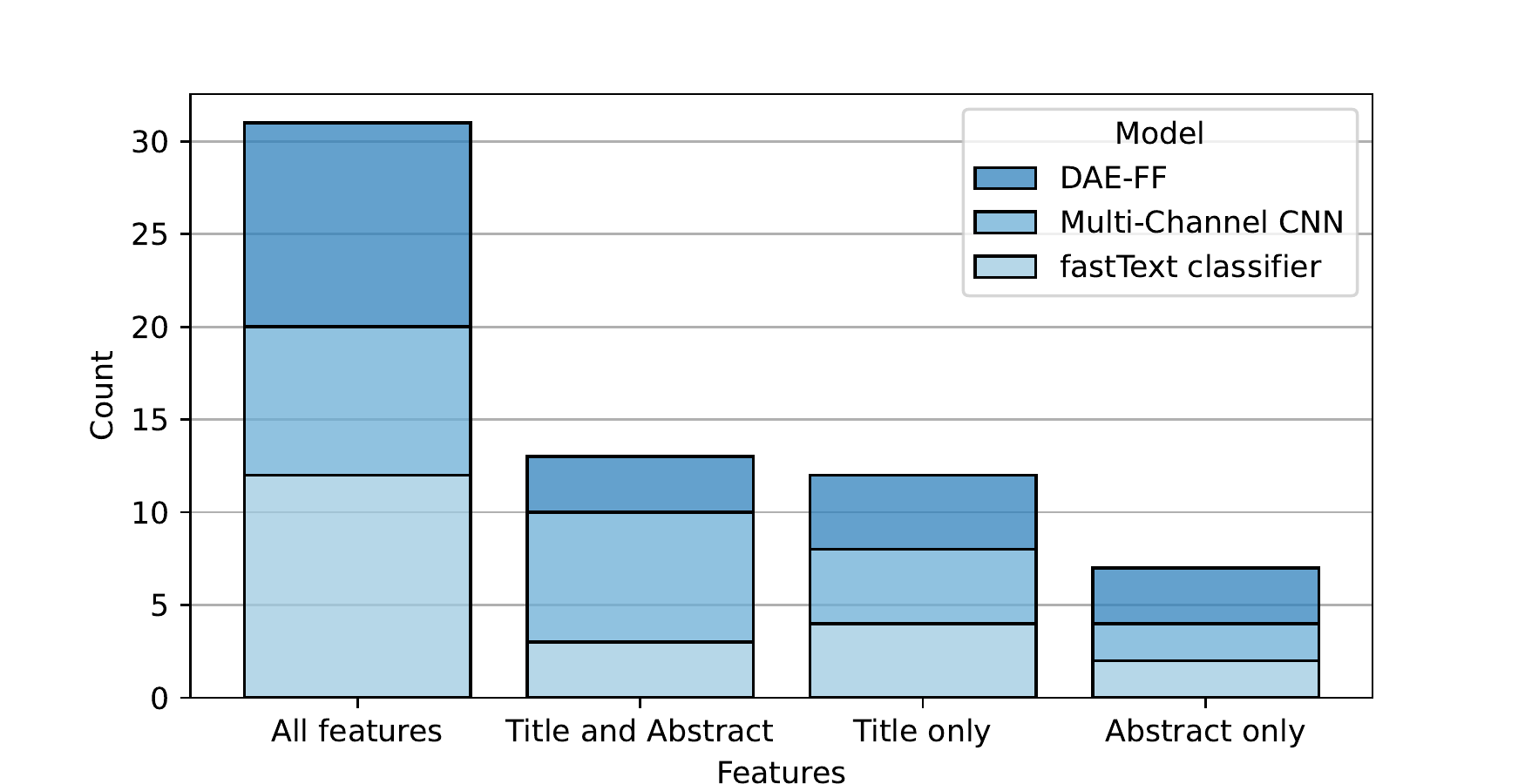}
  \caption{A count of experiments in which a model using a specific input feature achieved the best results. Models that use all available features scored the best results 49\% of times for a specific (model, dataset) combination.}
  \label{fig:features_sum}
\end{figure}

The best average WSS@95\% results are obtained for all three models when they use all available features (Figure \ref{fig:features_sum}). All models achieved better results when using just the abstract data compared to the titles alone. This reaffirms our common sense reasoning that titles alone are not sufficient for citation screening. However, there are some specific datasets for which best results were obtained when the input documents contained only titles or abstracts. While this experiment does not indicate why this is the case, we can offer some potential reasons: (1) it could be that eligible citations of these datasets are more similar in terms of titles or abstract; (2) it could be that these models are not able to retrieve relevant information when there is too much noise. Intra- and inter-class dataset similarity need to be further evaluated in future studies.

As presented in Table \ref{tab:wss95-results}, the fastText classifier model was not able to outperform the original results from Paper A and B. However, compared to our replicated results of Paper B, the fastText classifier achieves higher WSS@95\% scores on 18 out of 23 datasets. It is also more robust to random initialisation compared to Multi-Channel CNN.

\subsection{Training time}

\begin{figure}[!htb]

\centering
  \includegraphics[width=0.8\linewidth]{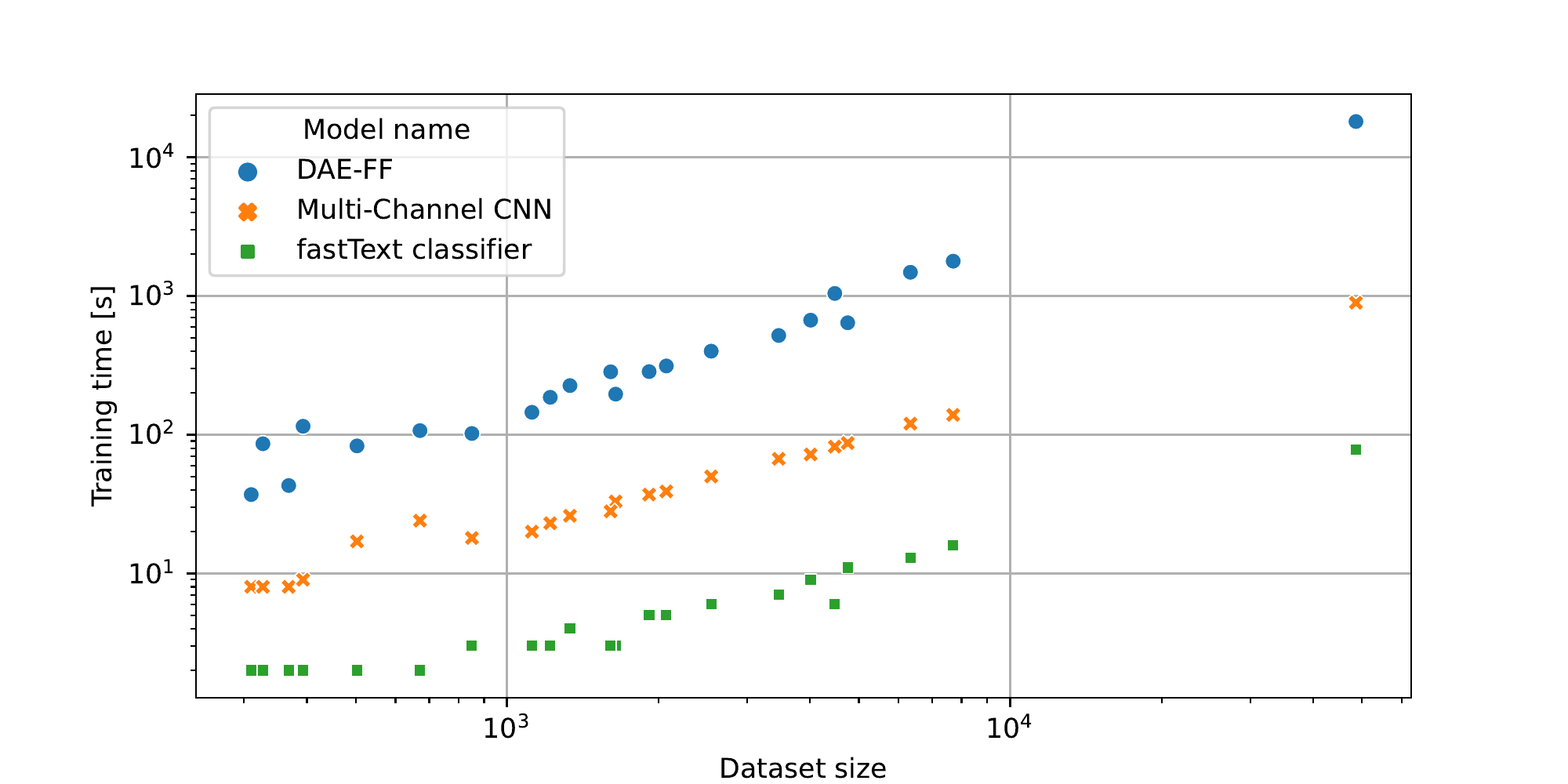}
  \caption{The relationship between dataset size and a model training time for the three evaluated models. Both training time and dataset size are shown on a logarithmic scale.}
  \label{fig:train_time}

\end{figure}

We computed the training time for each of the models. The relationship between dataset size and model training time is visualised in Figure \ref{fig:train_time}. For the DAE-FF model, we calculated both the training procedure of denoising autoencoder, feed-forward networks, and linear SVM. The DAE component is the most time-absorbing component as it consumes, on average, 93.5\% of the total training time. For the fastText and Multi-Channel CNN models, we calculated the training procedure of the binary classifier.

For small datasets containing less than 1000 documents, one validation fold for fastText took on average 2 seconds, for Multi-Channel CNN 13 seconds, and DAE-FF 82 seconds. Training time difference increases for larger models, where the speed of fastText is even more significant. For the largest dataset, \textit{Transgenerational}, the mean training time for fastText is 78 seconds, for Multi-Channel CNN 894 seconds and for DAE-FF, it is 18,108 seconds. On average, the fastText model is 72 times faster than DAE-FF and more than eight times faster than Multi-Channel CNN, although this dependency is not linear and favours fastText for larger datasets.

\subsection{Precision@95\%recall}

Finally, we measure the precision at a recall level of 95\%, a metric proposed by Paper A. Table \ref{tab:precision} shows mean scores for each model across all three review groups. Similarly to the WSS@95\% metric, the best performing model is DAE-FF achieving a mean precision@95\%recall on 21 datasets equal to 0.167. This method outperforms Multi-Channel CNN and fastText models by 3.2\% and 4.6\%, respectively. Paper A reported average precision@95\%recall equal to 19\% over 23 review datasets, which is comparable with our findings. Paper B does not report this score, so we cannot compare our results regarding the Multi-Channel CNN model.

\begin{table}[!ht]

\centering
\caption{Precision at 95\% recall results for the three models averaged on 21 benchmark datasets. We did not measure the precision for the two largest SWIFT datasets: \textit{Transgenerational} and \textit{Neuropathic pain}.}
\label{tab:precision}
\begin{tabular}{
>{\columncolor[HTML]{FFFFFF}}c |
>{\columncolor[HTML]{FFFFFF}}c |
>{\columncolor[HTML]{FFFFFF}}c |
>{\columncolor[HTML]{FFFFFF}}c }
\hline
                      & DAE-FF         & Multi-Channel CNN & fastText classifier \\ \hline
Average Drug          & \textbf{.143} & .121             & .112               \\ \hline
Average Clinical      & \textbf{.324} & .221             & .230               \\ \hline
Average SWIFT         & \textbf{.127} & .091             & .058               \\ \hline
Average (21 datasets) & \textbf{.167} & .135             & .121               \\ \hline
\end{tabular}
\end{table}

\section{Conclusions}

This work replicates two recent papers on automated citation screening for systematic literature reviews using deep neural networks. The model proposed by Paper A consists of a denoising autoencoder combined with feed-forward and SVM layers (DAE-FF). Paper B introduces a multi-channel convolutional neural network (Multi-Channel CNN). We used the 23 publicly available datasets to measure the quality of both models. The average delta between our replicated results and the original ones from Paper A is 3.59\%. Considering that we do not know the random seed used for the training of original models, we can conclude that the replication of Paper A was successful. The average delta for Paper B is 17.63\%. In addition to that, this model is characterised by a significant variance, so we cannot claim successful replication of this method.

Subsequently, we evaluated the fastText classifier and compared its performance to the replicated models. This shallow neural network model based on averaging word embeddings achieved better WSS@95\% results when compared to replicated scores from Paper B and, at the same time, is on average 72 and 8 times faster during training than both Paper A and B models.

None of the tested models can outperform all the others across all the datasets. DAE-FF achieves the best average results, though it is still worse when compared to a statistical method with the log-linear model. Models using all available features (title, abstract, author and journal information) perform best on the average of 21 datasets when compared to just using a title, abstract or both.

Availability of the code alone does not guarantee a replicable experimental setup. If the project was not documented for the specific software versions, it might be challenging to reconstruct these requirements based exclusively on the code, especially if the experiments were conducted some time ago. In the case of code written in Python, explicitly writing environment version with, for example, \textit{requirements.txt} or conda's \textit{environment.yml} files should be sufficient in most of the cases to save time for researchers trying to replicate the experiments.

\paragraph{Acknowledgements.} This work was supported by the EU Horizon 2020 ITN/ETN on Domain Specific Systems for Information Extraction and Retrieval -- DoSSIER (H2020-EU.1.3.1., ID: 860721).


\bibliographystyle{splncs04}
\bibliography{ref.bib}

\end{document}